\documentclass{vgtc}                          

\graphicspath{{figures/}{pictures/}{images/}{./}} 

\usepackage{times}                     

\usepackage{tabu}                      
\usepackage{booktabs}                  
\usepackage{lipsum}                    
\usepackage{mwe}        
\usepackage{amsmath}
\usepackage{amssymb}
\usepackage{adjustbox}

\usepackage{mathptmx}                  

\onlineid{3203}

\vgtccategory{Application}

\vgtcinsertpkg

\title{VAIR: Visual Analytics for Injury Risk Exploration in Sports}

\author{Chunggi Lee\thanks{e-mail: chunggi\_lee@g.harvard.edu}\\ %
        \scriptsize Harvard University %
\and Ut Gong\\ %
     \scriptsize Harvard University %
\and Tica Lin\\ %
     \scriptsize Dolby Laboratories %
\and Stefanie Zollmann \\ %
     \scriptsize Aarhus University %
\and Scott A Epsley \\ %
     \scriptsize Gotham FC %
\and Adam Petway \\ %
     \scriptsize University of Louisville %
\and Hanspeter Pfister\thanks{e-mail: pfister@g.harvard.edu}\\ %
    {\scriptsize Harvard University}}

\teaser{
  \centering
  \begin{adjustbox}{center, max width=0.265\linewidth}
    \includegraphics{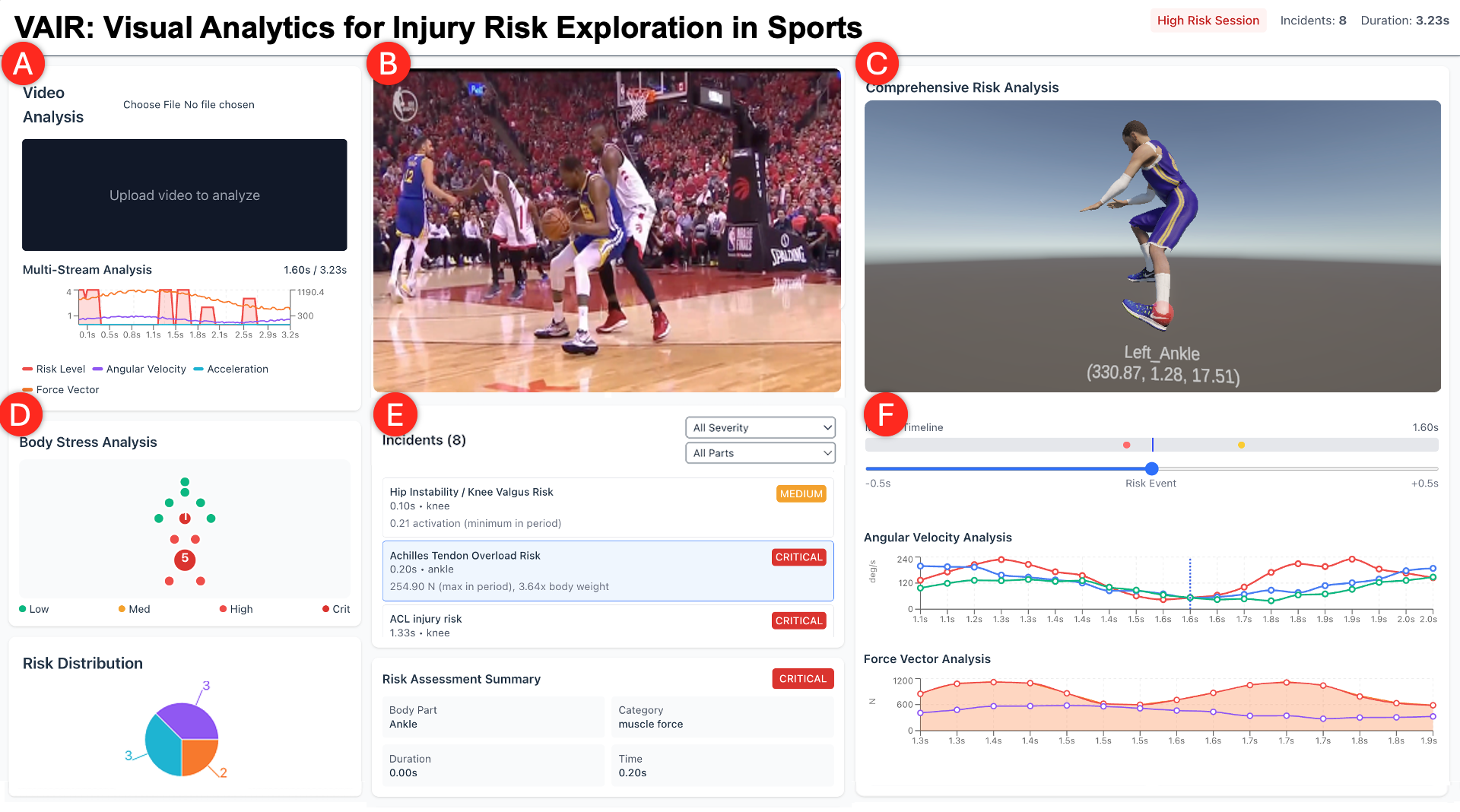}
  \end{adjustbox}
  \caption{VAIR system interface showing multi-perspective injury risk analysis for a basketball motion clip. The interface integrates synchronized video playback, 3D reconstruction with risk annotations, multi-variate biomechanical data line charts, and spatial stress summaries. Users can explore high-risk segments, examine joint-level dynamics, and interpret injury mechanisms in the context of real gameplay.}
  \label{fig:teaser}
}

\abstract{
    Injury prevention in sports requires understanding how biomechanical risks emerge from movement patterns captured in real-world scenarios. However, identifying and interpreting injury-prone events from raw video remains difficult and time-consuming. We present \textbf{VAIR}, a visual analytics system that supports injury risk analysis using 3D human motion reconstructed from sports video. VAIR combines pose estimation, biomechanical simulation, and synchronized visualizations to help users explore how joint-level risk indicators evolve over time. Domain experts can inspect movement segments through temporally aligned joint angles, angular velocity, and internal forces to detect patterns associated with known injury mechanisms. Through case studies involving Achilles tendon and Anterior cruciate ligament (ACL) injuries in basketball, we show that VAIR enables more efficient identification and interpretation of risky movements. Expert feedback confirms that VAIR improves diagnostic reasoning and supports both retrospective analysis and proactive intervention planning.
} 


\begin{document}



\newcommand{\name}{VAIR\space}

\maketitle
\section{Introduction} 

Sports injuries are a major barrier to sustained physical activity and present a serious public health concern, contributing to increased healthcare costs and reduced quality of life \cite{bell2019public, warburton2006health, haskell2007physical}. For decades, researchers have sought to understand injury mechanisms and develop prevention strategies \cite{defroda2016two, uhlrich2023opencap, parkkari2001possible}. Many existing approaches \cite{rebelo2023data, arzehgar2025sensor} focus on pre-event interventions that target individual athletes such as strength training, wearable sensors, and motion capture in controlled laboratory settings. These efforts have deepened our understanding of biomechanical risk factors like joint load, movement asymmetry, and neuromuscular control.

While these methods are capable of capturing precise biomechanical data, they are not easily scalable to real-world sports environments where athletes move freely and external conditions are constantly changing. This limits their practical applicability for coaches, trainers, and clinicians who need solutions that function reliably in naturalistic, high-variability scenarios without extensive setup. Even when biomechanical data is collected, it is often presented in a way that does not directly support decision-making. The results are typically reported as isolated numerical values (e.g., joint torque, angular velocity, or trajectory deviations) without connecting them to broader patterns of risk or situating them within the context of injury mechanisms. In many cases \cite{defroda2016two}, visualizations simply overlay these values on video footage, offering surface-level illustrations rather than structured insights. These overlays might show where movement occurs but fail to explain why the movement poses a risk or how it could be mitigated. This lack of integration makes it difficult for domain experts to translate biomechanical findings into actionable strategies. 

To address these limitations, we present VAIR, a visual analytics system designed to support contextualized injury risk assessment using 3D human motion reconstructed from video. VAIR enables sports scientists, clinicians, and biomechanical researchers to explore how injury risks evolve over time and compare joint-level biomechanical characteristics across individuals or sessions. By leveraging pose estimation and biomechanical simulation, VAIR bridges the gap between observational analysis and quantitative assessment. Users can examine temporally aligned biomechanical signals, such as joint angles, angular velocity, and internal forces, to detect risky movement patterns that may contribute to injury.

We evaluated VAIR through case studies with domain experts, including team physicians and rehabilitation specialists, using real-world basketball footage. Experts used the system to analyze injury-prone movements such as cutting, landing, and player collisions. The analysis included both Achilles tendon and ACL injury scenarios. Feedback indicated that the system's synchronized visualizations helped identify high-risk moments and supported detailed biomechanical interpretation. Experts emphasized the value of isolating movement segments and inspecting joint-level indicators to understand injury mechanisms and develop informed intervention strategies.

In summary, our contributions are threefold:
(1) We develop a visual analytics system that integrates 3D motion reconstruction, biomechanical simulation, and multivariate feature analysis;
(2) We propose a workflow that links low-level motion data with interpretable risk indicators;
(3) We validate the system through expert case studies involving Achilles tendon and ACL injuries, demonstrating improved reasoning and contextual understanding of real-world injury risks.

\section{Related Work}

\subsection{Biomechanical Injury Analysis}
Understanding the biomechanical origins of sports injuries has long relied on laboratory-based methods such as motion capture, force plates, and wearable sensors. These methods offer precise kinematic and kinetic data but are limited in ecological validity and scalability \cite{rebelo2023data, arzehgar2025sensor}.
However, wearing such sensors during actual competitive matches is often impractical, which makes it difficult to capture biomechanical signals at the moment injuries occur. This gap significantly limits our ability to study real-world injury mechanisms as they unfold in context.
In contrast, our approach leverages only in-game video footage to reconstruct 3D human motion, enabling injury analysis in ecologically valid environments without the need for intrusive instrumentation.

\subsection{Video-based 3D Pose Reconstruction}
Recent advances in human mesh recovery (e.g., Skinned Multi-Person Linear model (SMPL) \cite{loper2023smpl}, VIBE \cite{kocabas2020vibe}, and Co-Motion \cite{newell2025comotion}) enable the estimation of 3D body motion from monocular videos. Co-Motion proposed SMPL-based 3D human mesh models to enable both motion reconstruction and multi-person tracking from video, providing temporally consistent joint trajectories in unconstrained environments. By leveraging SMPL's parametric representation, these approaches estimate human pose and shape across time, enabling downstream biomechanical analysis without the need for marker-based motion capture systems. Although such methods do not yet achieve the precision of sensor-based instrumentation, they offer broader applicability across diverse real-world settings. By enabling markerless reconstruction and tracking, they establish a foundation for delivering visual feedback and facilitating injury risk assessment.

\subsection{Visual Analytics for Movement and Risk}
Visual analytics (VA) has been explored for movement assessment in domains such as rehabilitation and sports. VA systems such as PoseCoach \cite{liu2022posecoach, palmas2014movexp} enable users to analyze joint trajectories or muscle activity to aid interpretation. While these systems provide useful insights, they typically focus on raw data presentation or event detection without linking biomechanical signals to broader injury mechanisms or contextual understanding. In practice, many domain experts still rely on manual video annotation tools such as Dartfish\cite{dartfish} to visually overlay biomechanical cues and assess injury-prone movements frame by frame, which is time-consuming and lacks integration with automated analysis pipelines. This highlights a gap between automated VA systems and the practical, context-aware workflows of clinicians and coaches.
Our work bridges this gap by integrating 3D pose reconstruction and biomechanical simulation, and by structuring risk analysis to support contextualized and interpretable injury assessment grounded in real-world video footage.

\begin{figure*}[t]
    \centering
    \includegraphics[width=\textwidth]{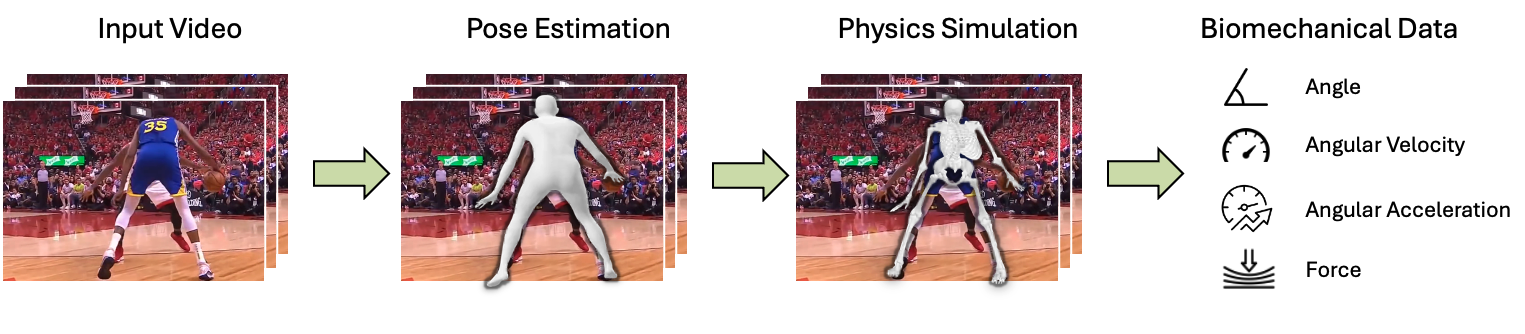}
    \caption{Overview of our pipeline. From an input video, we reconstruct 3D human pose using a tracking and reconstruction model, simulate joint dynamics with a physics-based simulation model, and extract biomechanical quantities such as angle, angular velocity, angular acceleration, and force.}
    \label{fig:pipeline}
\end{figure*}

\section{Design Requirements}
To support effective injury risk analysis in real-world environments, we collaborated with two domain experts to derive the following design requirements for our system. One expert is a sports physical therapist with over 25 years of experience in elite sports medicine in the NBA, NFL, NHL, MLB. His clinical background in injury rehabilitation and diagnostic imaging provided essential perspectives on real-world injury mechanisms and medical workflows. The other expert is a strength and conditioning specialist with over a decade of experience in the NBA and NCAA. He has served in performance and biomechanical roles for teams and holds a doctorate in sports science with research focused on neuromuscular profiling in elite basketball athletes.  Together, these domain experts participated in a formative study involving iterative discussions, task analyses, and informal interviews. Their insights directly shaped the design requirements of our system, grounded in the practical needs of injury risk assessment.

\textbf{R1. 3D Motion Reconstruction for Multi-Angle Movement Analysis.}
To understand the causes and risks of sports injuries, it is necessary to analyze specific scenes or athlete movements from video footage. However, traditional 2D video lacks the spatial information needed to observe movement from multiple perspectives. The system should therefore support the reconstruction of 3D human motion and body structure from video recordings. This functionality can greatly assist biomechanical analysis in real-world settings by reducing reliance on specialized hardware such as motion capture systems or wearable sensors. A reconstructed 3D view enables flexible observation of movement from arbitrary angles (e.g., side, top, free 3D view) \cite{uhlrich2023opencap}, enabling users to examine motion patterns that are difficult to perceive from fixed camera viewpoints. The reconstructed data should also be temporally consistent and spatially accurate to support further estimation of features such as joint velocity, angular velocity, and acceleration. Prior work \cite{defroda2016two} has shown that 3D motion estimation from video provides access to key biomechanical measurements such as joint angles and segment velocities, using affordable and widely available tools without the need for specialized lab environments. These observations highlight the need for a system that can reconstruct accurate and consistent 3D human motion from video, enabling contextualized and accessible injury risk analysis in real-world scenarios.

\textbf{R2. Multivariate Biomechanical Motion Data Visualization.}
Injury mechanisms emerge through biomechanical motion data such as joint angles, torques, and forces. To capture these critical indicators, the system should enable users to explore a variety of biomechanical features (e.g., joint velocity, angular velocity, torque, forces) across different anatomical regions (e.g., ankle, knee, and hip).
To support effective interpretation, the system should visualize these features through time-aligned plots (e.g., angular velocity and force vector changes), schematic body maps indicating joint-level stress severity, and structured summaries of detected risk events categorized by body region and severity. Such visual representations are essential for revealing temporal trends, localized stress patterns, and abnormal motion characteristics that may not be apparent in raw numerical data alone. These components collectively enable users to identify injury-prone moments and understand the underlying biomechanical context.

\textbf{R3. Estimation and Representation of Biomechanical Risk Indicators.}
While multi-feature visualization can reveal suspicious patterns, clear estimation of biomechanical risk indicators is necessary to confirm and quantify potential injuries. To support injury risk analysis, it is essential to estimate biomechanical indicators (e.g., joint displacement, angular velocity, and segment acceleration) from reconstructed motion data in a temporally consistent and anatomically accurate manner. The system provides risk predictions derived from these biomechanical indicators, offering users a summarized view of potential injury risks. Based on this overview, users can conduct deeper analysis using multi-variate biomechanical motion data visualizations such as angular velocity and force patterns. This feature enable users to investigate the underlying causes and temporal characteristics of risky movements.

\section{Biomechanical Reconstruction and Simulation}

To realize the aforementioned design requirements, our system builds upon two key technical components. ~\autoref{fig:pipeline} provides an overview of our 3D reconstruction and simulation pipeline. Starting from a standard input video, the system first performs 3D pose estimation to reconstruct a temporally consistent human mesh using the pose tracking model \cite{newell2025comotion}. In our setting, we analyzed injury clips that were approximately 3–10 seconds long, and the reconstruction process typically completed within 30 seconds per video. This reconstructed motion is then used as input for physics-based simulation in OpenSim\cite{delp2007opensim}, which estimates internal biomechanical quantities such as joint angles, angular velocity, angular acceleration, and force. These computed indicators form the foundation for subsequent injury risk analysis, enabling the system to move beyond visual appearance and surface-level motion toward quantitative.

\subsection{3D Human Mesh Reconstruction from a Monocular Video}

To perform biomechanical analysis from monocular video input, we first require a reliable 3D representation of the human body that preserves anatomical structure and supports accurate motion estimation over time. We utilize Co-Motion~\cite{newell2025comotion}, a recent method that builds on the SMPL (Skinned Multi-Person Linear) model~\cite{loper2023smpl} to jointly reconstruct 3D human meshes and perform multi-person tracking from video. Co-Motion estimates temporally consistent SMPL parameters (e.g., both shape and pose) for each person in the scene. These meshes capture detailed surface geometry and support joint-level and segment-level analysis. By leveraging the parametric structure of SMPL, Co-Motion produces human body representations that are deformable, anatomically interpretable, and suitable for downstream biomechanical computation. The resulting 3D meshes, reconstructed directly from unconstrained video, provide a robust foundation for analyzing complex human motion in real-world scenarios without requiring specialized motion capture equipment.


\subsection{Multi-variate Biomechanical Motion Data Estimation}

To analyze injury risk and extract meaningful biomechanical motion data, it is necessary to estimate internal physical quantities such as torque and force. These indicators provide critical insights into how the body responds to dynamic movements and where stress accumulates. OpenSim~\cite{delp2007opensim} is a musculoskeletal modeling platform that enables the simulation of internal biomechanical forces based on motion kinematics. After reconstructing joint trajectories from the meshes, we convert them into biomechanical motion data by using OpenSim. These simulations enable us to estimate biomechanical indicators such as joint loading and torque, muscle activation asymmetries, ground reaction force patterns, rapid angular changes or asymmetries linked to injury risk. By integrating SMPL-based 3D reconstruction with OpenSim simulation, our system can visualize biomechanical stress across joints, detect abnormal movement patterns, and support early detection of injury risks.

\subsection{Biomechanical Risk Estimation}
\begin{figure*}[t]
    \centering
    \includegraphics[width=0.8\textwidth]{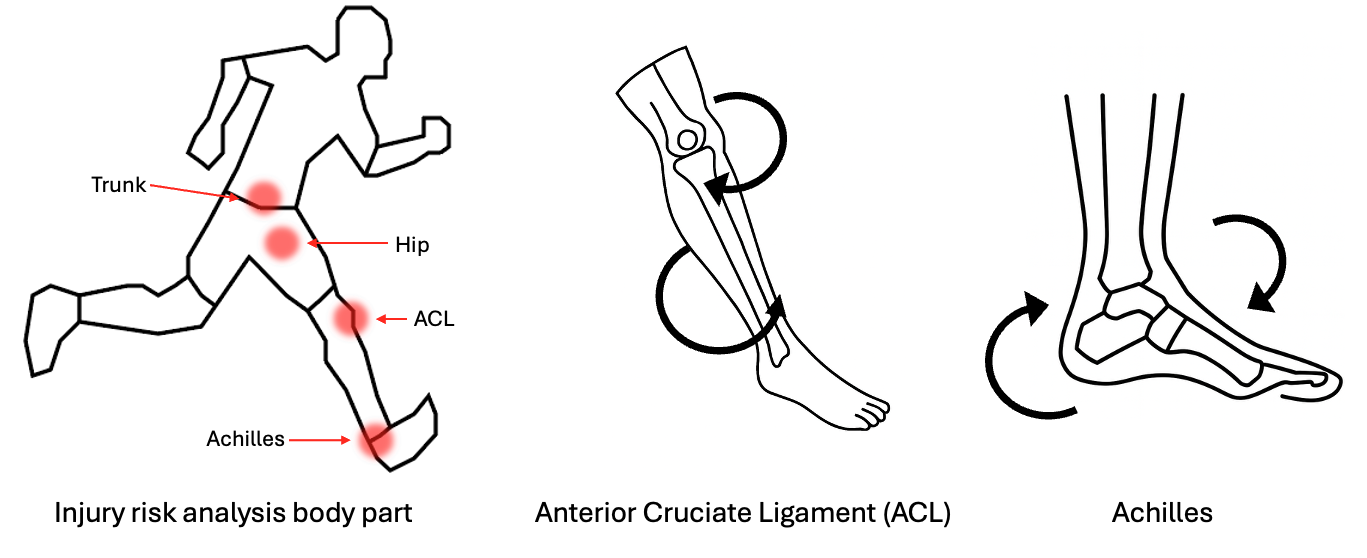}
    \caption{Key anatomical regions involved in injury risk analysis. The illustration highlights commonly affected body parts during athletic motion: (left) trunk, hip, ACL, and Achilles regions with red markers; (middle) anterior cruciate ligament (ACL) and its rotational vulnerability; and (right) Achilles tendon and associated foot joint stress patterns. These regions are commonly analyzed in biomechanical and injury-risk studies.}
    \label{fig:injuries}
\end{figure*}

As illustrated in ~\autoref{fig:injuries} (left), regions such as the trunk, hip, knee (particularly the ACL), and Achilles tendon are commonly involved in sports-related injuries due to their biomechanical role in load absorption and movement coordination. These regions are responsible for absorbing external loads, generating joint torque, and stabilizing body segments through coordinated angular motion. In our system, we analyze joint angles, joint torques, and  forces to detect abnormal biomechanical patterns that indicate elevated injury risk. We refer to prior literature~\cite{ge2025exploring, feeley2019can, zeitlin2023key, ekegren2009reliability, hewett2011mechanistic} to estimate injury risk based on our reconstructed motion data. By comparing the reconstructed joint angles and forces (e.g., knee, hip, trunk) to these known risk ranges, we identify frames and movements that may involve a higher chance of injury. This approach enables us to combine literature-based evidence with simulated kinematics to assess injury risk more systematically.

\subsubsection{Joint Angles}
Joint kinematics are foundational for understanding how specific movement strategies influence injury potential. Studies ~\cite{feeley2019can, ekegren2009reliability, della2022video} have examined joint angles to reveal how factors such as foot orientation or knee valgus relate to injury mechanisms. We directly utilized joint angles derived from SMPL pose parameters in Unity. 
Specifically, we used findings from~\cite{della2022video}, which report that ACL are more likely to occur when ankle dorsiflexion dorsiflexion (upward foot movement toward the shin) reaches up to 40° and knee flexion is approximately 22.5°, to define risk conditions in our system. By comparing reconstructed ankle and knee angles to these values, we identified frames in which the joints were placed in vulnerable configurations, providing users with interpretable cues of potential Achilles injury risk. The anterior cruciate ligament (ACL) is known to bear significant anterior restraining loads during knee flexion between 30° and 90°, particularly when internal rotation and valgus forces are present~\cite{dargel2007biomechanics}. By comparing reconstructed knee angles to a threshold, we indicated potential injury risk.
~\autoref{fig:injuries} (center) illustrates how excessive rotation or limited knee flexion can stress the ACL, with arrows showing high-risk joint movements.

\subsubsection{Joint Contact Forces}
Joint contact forces, such as the pressure between bones in the knee or the front-to-back pulling forces, are important.
They help us understand how much stress occurs inside the joint, beyond what we can observe from motion alone. We utilized this model \cite{addbiomechanics2024} to estimate these internal forces based on movement and force data. These internal loading patterns are further illustrated in ~\autoref{fig:injuries} (right). The center diagram shows the rotational loading of the ankle, which affects Achilles tendon stress.

\begin{figure*}[t]
    \centering
    \includegraphics[width=\textwidth]{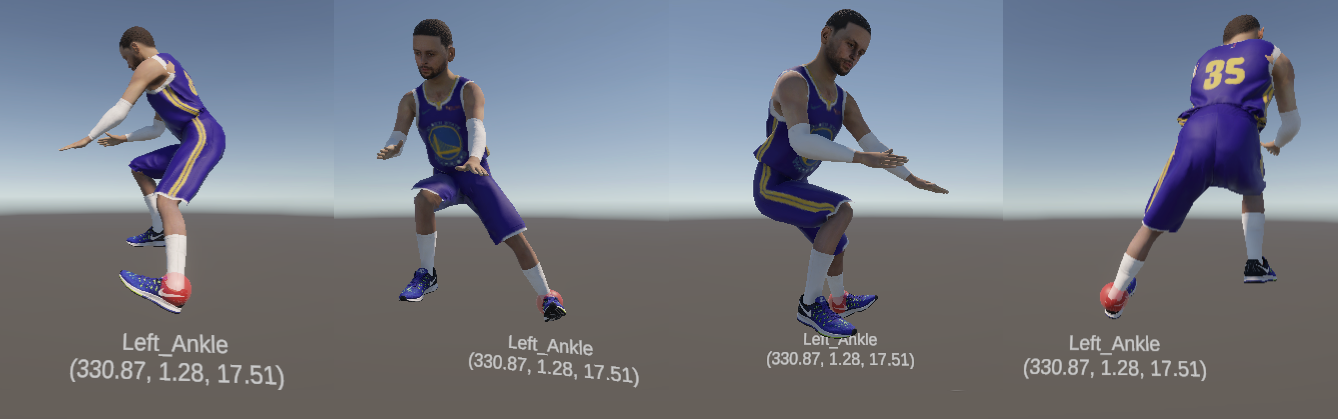}
    \caption{ Multi-angle visualization of reconstructed offensive drive motion. By changing viewpoints, analysts can examine spatial dynamics such as foot placement and knee rotation to Achilles tendon risk. The annotated values represent the left ankle’s rotation in Euler angles (degrees), highlighting joint orientation during the movement.}
    \label{fig:comparison}
    \vspace{-0.1cm}
\end{figure*}

\section{VAIR System}

To support multi-perspective injury risk analysis from video-based human motion data, we developed the \name. As illustrated in ~\autoref{fig:teaser}, the system offers multi-coordinated views that support exploration from raw motion input to multi-perspective and multi-variate biomechanical motion data analysis and estimating risks. 

\subsection{3D Motion Reconstruction for Multi-Angle Movement Analysis (R1)}
The analysis begins in the \textbf{Video Analysis} panel, where users upload a 2D video as shown in \autoref{fig:teaser} (A), showing the original video in \autoref{fig:teaser} (B). The system reconstructs 3D human motion using pose estimation pipelines and generates temporally consistent skeletal sequences. This reconstruction enables users to explore movement from arbitrary viewpoints (e.g., frontal, lateral, top-down) in \autoref{fig:teaser} (C), overcoming the limitations of single-camera footage. Users can observe spatial motions that are critical to understanding injury-prone behaviors, such as misalignment during landing or rotational asymmetries during cutting maneuvers. 

\subsection{Multi-Feature Biomechanical Visualization (R2)}
After reconstruction, biomechanical signals such as joint angles and force vectors are computed.
The \textbf{Body Stress Analysis} panel in \autoref{fig:teaser} provides insights into joint-specific stress levels, with severity encoded by color. This panel represents potential risks from simulated models. In addition, \textbf{Risk Distribution} shows the risk distribution of biomechanical stress across the body. This spatial visualization supports identification of localized mechanical stress and helps interpret motion anomalies beyond raw signal plots.
In addition, these features are shown in the \textbf{Multi-Stream Analysis}, \textbf{Angular Velocity Analysis}, and \textbf{Force Vector Analysis} panels in in \autoref{fig:teaser} (A and F). Each plot is temporally aligned, enabling users to detect rapid spikes or abnormal trajectories that may signal biomechanical overload or instability. Multi-Stream Analysis demonstrates the summary view of biomechanical data. This time-series data is further contextualized through the Angular Velocity Analysis and Force Vector Analysis. \name enables user to explore the deeper analysis of each individual body parts.

\subsection{Risk Estimation and Event-Based Summarization (R3)}
To enhance interpretability and focus user attention, the system generates risk indicators using rule-based and data-driven thresholds from previous literature~\cite{ge2025exploring, feeley2019can, zeitlin2023key, ekegren2009reliability, hewett2011mechanistic}. Detected incidents are listed in the \textbf{Incidents} panel (~\autoref{fig:teaser} (E), where each entry includes time of occurrence, anatomical location, and severity levels. These events are summarized in the \textbf{Risk Assessment Summary}, which offers an overview of critical occurrences per session. Risk indicators are also encoded in the \textbf{Comprehensive Risk Analysis} panel via a timeline of incident markers aligned with 3D animation. This facilitates event-driven review with precise spatiotemporal context.

\section{Case Study: Achilles and ACL Injury Risk Anaysis}

\begin{figure*}[t]
    \centering
    \includegraphics[width=\textwidth]{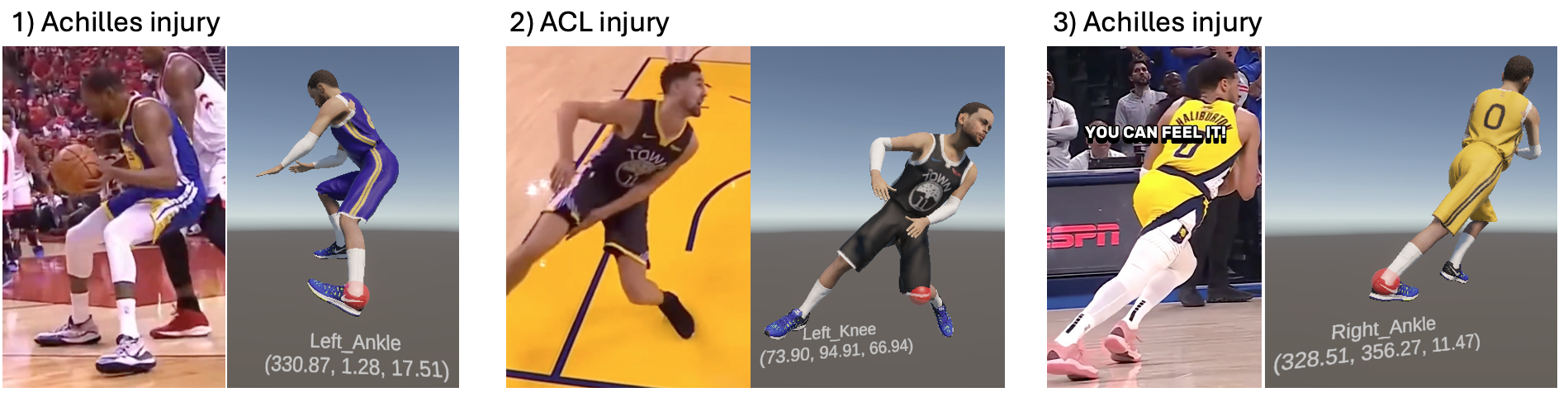}
    \caption{Representative reconstructions across injury-prone movements. Our pipeline generalizes to various motions including jump landings and player collisions, supporting broader biomechanical injury analysis.}
    \label{fig:injury_example}
    \vspace{-0.5cm}
\end{figure*}

To demonstrate the practical utility of our system in identifying injury risks, we present a case study involving two doctors (P1 and P2) from the basketball team and one therapist (P3) analyzing an three different injury motions. We used three different  video clips from NBA games \cite{youtube_kbzZbQR0IY, youtube_QZuWch0CP70, youtube_ZnSEgiRbHs}. 
During the study, P1 and P3 collaboratively interacted with the VAIR system, while P2 conducted the analysis independently.
First of all, we analyze one particular injury video \cite{youtube_kbzZbQR0IY}, which captures an Achilles injury that occurred during an offensive play. As illustrated in ~\autoref{fig:teaser} (C), the player plants the left foot while leaning forward, resulting in an ankle dorsiflexion angle of approximately -29°. This posture represents a high-risk condition, consistent with the injury-prone patterns described in Section 4.3.1. This movement pattern is biomechanically demanding and places the Achilles tendon at elevated risk, as it involves forward leaning, deep dorsiflexion, and knee rotation, all of which contribute to high eccentric loading.

In the \textbf{Multi\-Stream Feature Visualization} panel (A), users can examine multiple time-synchronized signals such as angular velocity, acceleration, and external force magnitude. Around the moment of foot planting, notable peaks in plantarflexion torque and trunk rotation velocity are observed. In the \textbf{Incident Detection and Summary} panel (E), the system automatically flags these frames as high-risk events on Achilles tendon stress. For instance, one of the flagged incidents indicates a force magnitude of 254.9N applied to the ankle—equivalent to over 3.6 times body weight—alongside rotational movement. In the \textbf{Body Stress Analysis} (D) and \textbf{Risk Distribution} panels, this risk is visualized spatially, showing concentrated load on the left ankle region. This case illustrates how our system integrates multi-modal signals and biomechanical reasoning to enable domain experts to detect, interpret, and contextualize high-risk movements directly from real-world game footage.

\subsection{Multi-View Risk Analysis and Comparative Injury Scenario Analysis}

To further support detailed biomechanical interpretation, ~\autoref{fig:comparison} visualizes the same motion from four distinct viewpoints. These multi-angle renderings enable domain experts to observe subtle spatial dynamics (e.g., ankle and knee rotation) that are not easily captured from a single camera perspective. Such variations in viewpoint are crucial for interpreting injury-relevant kinematic features, especially in movements involving rotation or eccentric loading, which are commonly associated with Achilles tendon stress. In addition, our pipeline generalizes beyond a single movement type. As shown in~\autoref{fig:injury_example}, we apply our system to a diverse set of real-world injury scenarios, including both Achilles and ACL injuries. By reconstructing these motions into anatomically grounded 3D representations, the system enables comparative analysis of injury risk across different biomechanical contexts. 

In the second example, the motion illustrates an ACL injury scenario. Observational risk screening literature supports—dynamic knee valgus, defined as medial collapse of the knee during hip and knee flexion—is a known risk factor for ACL injuries.
Reflecting this, our reconstructed pose shows the left knee with approximate angles: flexion $\sim$74°, abduction $\sim$95°, and internal rotation  $\sim$67°. Although the knee flexion itself exceeds the “ \textless30°” threshold reported in~\cite{dargel2007biomechanics}, the excessive abduction and internal rotation clearly align with the valgus collapse mechanics identified in these studies as critical contributors to ACL injury risk.

The third example again involves an Achilles injury, this time on the right foot, with pronounced plantar flexion ($\sim$-32°) and knee rotation, a movement known to produce high eccentric loading. These examples demonstrate the versatility of our approach in handling varied movement patterns and support broader generalization across different injury contexts in sports biomechanics. Collectively, ~\autoref{fig:comparison} and~\autoref{fig:injury_example} illustrate not only the system's capacity to support fine-grained visual inspection, but also its applicability to a broad range of injury-prone actions. This enables analysts to compare movement patterns across scenarios, enhancing their ability to identify risky conditions with both anatomical and situational awareness.

Currently, injury risk analysis in elite sports settings is largely manual: experts must first identify potential injury clips by reviewing hours of footage, then use commercial tools \cite{dartfish} to manually annotate joint positions and measure angles for biomechanical interpretation. This process is time-consuming and inconsistent across evaluators. Our system streamlines this workflow by automating 3D motion reconstruction and feature extraction from standard video, reducing expert burden while increasing scalability and reproducibility. Furthermore, the system lays the foundation for cumulative stress tracking over time, which could support tailored training loads, individualized rest schedules, and early intervention strategies.

\section{User Feedback \& Future Work}

During our case study and discussions with domain experts, three key areas of user needs emerged:

\subsection{Discovering Injury-Prone Patterns Across Video}

Users appreciated that the system significantly reduces the burden of manual annotation. As P1 noted, \textit{"It's much better than manually annotating and analyzing every frame."} By enabling time-based navigation through biomechanical signal peaks and providing visual cues like joint angles, the system supports faster and more efficient review of movement data. However, they also emphasized that identifying potential injury-prone segments remains a challenge. P3 mentioned \textit{"It's still hard to find parts in the video where an injury is likely to occur."} One reason for this limitation is that our current examples focus on single-person scenarios with known injury events. In contrast, real-world use cases often involve continuous footage without labeled injury moments, and may include multiple individuals moving simultaneously. This makes it difficult to detect subtle precursors to injury or distinguish which subject is at risk. To address this, we plan to integrate event-level segmentation algorithms that automatically flag high-risk segments based on continuous biomechanical patterns. These will be displayed in a fragmented visualization format with risk annotations and color-coded cues, enabling users to quickly triage large video collections and focus on areas of concern.

\subsection{Understanding Player Contacts and Interactions}

Injuries result not only from an individual’s movement but also from interactions with other players, such as physical contact, balance disruption, or defensive pressure. Users noted that the current system focuses primarily on individual kinematics and force profiles. There was a clear need for features that visualize proximity, collisions, or joint perturbation caused by opponent contact, especially in competitive sports settings like basketball. P1 said \textit{"Understanding whether an injury is contact or non-contact is critical, as it fundamentally changes how we interpret its cause and prevention."} We aim to incorporate multi-person tracking to detect and visualize physical interactions between players. By analyzing joint proximity, contact surfaces, and reactive movements, the system can reveal external factors contributing to injury mechanisms. These insights can enrich the incident analysis panel and better reflect real-world injury causation.

\subsection{Spatial Context and Court-Level Localization}

Experts highlighted the importance of spatial awareness—knowing where on the court an injury or risky maneuver occurred. This spatial grounding helps contextualize biomechanical risk in terms of game strategy (e.g., close to the basket, on the perimeter) and environmental conditions (e.g., surface type, defender location). Users requested court overlays or visual annotations to integrate positional context into the biomechanical analysis. We also plan to align motion trajectories with a court model or spatial map, enabling spatially anchored visualization. This would enable users to view biomechanical stress alongside tactical positioning (e.g., pick-and-rolls, isolation plays), bridging the gap between movement quality and game context.

\subsection{Integration with XR for Immersive Risk Review}

To further enhance situational understanding, we plan to explore integration with Extended Reality (XR) platforms. By projecting reconstructed player movements and force vectors into immersive 3D environments, XR can enable embodied review of injury-prone actions. Coaches and analysts could walk around the scene, switch perspectives, or experience risky maneuvers from the player’s point of view, facilitating intuitive reasoning about injury mechanics. As P2 suggested, such embodied review could provide a powerful tool for both injury prevention and education in applied sports settings.

This XR-based review is not limited to basketball or a specific type of motion. It can be extended to a wide range of sports and movement types such as pivoting in soccer, landing in volleyball, or sudden deceleration in football. Biomechanical risk is influenced by complex spatial-temporal dynamics. By generalizing the pipeline to accommodate different motion data inputs, we aim to support broader applications in sports science, rehabilitation, and return-to-play assessment.

\subsection{Lack of Objective Performance Evaluation}
While our system benefits from recent advances in video-based 3D human pose estimation, certain limitations remain. When video quality is high and subjects are clearly visible, recent models \cite{newell2025comotion, sarandi2024neural} can produce results comparable to those of marker-based motion capture systems. However, errors still occur in cases involving occlusion \cite{goel2023humans}, such as overlapping players with most errors occurring in the limbs, or when subjects appear small or partially out of frame. These limitations may result in incorrect joint reconstruction or inaccurate biomechanical feature estimation. The current evaluation lacks quantitative metrics such as pose estimation accuracy, tracking performance. A more rigorous assessment with pose estimation remains an important direction for future work.

\section{Conclusion}
We introduced VAIR, a visual analytics system for injury risk analysis via video-based 3D motion reconstruction and biomechanical simulation. Case studies demonstrated its effectiveness in interpreting high-risk movements using joint-level indicators. While useful for single-person injuries, feedback revealed limitations in multi-person scenarios and continuous video. To improve real-world applicability, we plan to add automated risk detection, contact analysis, court localization, and XR-based immersive review.

\acknowledgments{
We thank the anonymous reviewers for their constructive feedback, which helped improve the quality of this paper.
}

\bibliographystyle{abbrv-doi}

\bibliography{template}
\end{document}